\documentclass[aps,prc,twocolumn,floatfix,superscriptaddress]{revtex4-1}  
\usepackage{amsmath}
\usepackage{amsfonts}
\usepackage{amssymb}
\usepackage{graphicx}
\usepackage{epstopdf}
\usepackage{xcolor}
\usepackage{hyperref}

\begin{document}
\title{
Asymptotic normalization coefficients of alpha-particle removal from 
$^{16}$O($3^-,2^+,1^-$)}

\author{L. D. Blokhintsev}
\affiliation{Skobeltsyn Institute of Nuclear Physics, Lomonosov Moscow State University,  Moscow 119991, Russia}

\author{A. S. Kadyrov}
\affiliation{Department of Physics and Astronomy and Curtin Institute for Computation, Curtin University, GPO Box U1987, Perth, WA 6845, Australia}

\author{A. M. Mukhamedzhanov}
\affiliation{Cyclotron Institute, Texas A\&M University, College Station, TX 77843, USA}

\author{D. A. Savin}
\affiliation{Skobeltsyn Institute of Nuclear Physics, Lomonosov Moscow State University,  Moscow 119991, Russia}

\begin{abstract}
Asymptotic normalization coefficients (ANC)  determine the overall normalization of cross sections of peripheral radiative  capture reactions.
In  a recent paper [Blokhintsev {\em et al.}, Eur. Phys. J. A {\bf 58}, 257 (2022)],
we considered the ANC $C_0$ for the virtual decay $^{16}$O$(0^+; 6.05$ MeV)$\to \alpha+^{12}$C(g.s.). In the present paper, which can be regarded as a  continuation of the
previous, we treat the ANCs $C_l$ for the vertices $^{16}$O$(J^\pi)\to \alpha+^{12}$C(g.s.) corresponding to  the other three bound excited states of $^{16}$O ($J^\pi=3^-$, $2^+$, $1^-$, $l=J$).
 ANCs $C_l$ ($l=3,\,2,\,1$)
 are found by analytic continuation in energy of the $\alpha^{12}$C $l$-wave partial scattering amplitudes, known from the phase-shift analysis of experimental data,  to the pole corresponding to the 
$^{16}$O bound state and lying in the unphysical region of negative energies. 
To determine $C_l$, the  scattering data are approximated by the sum of polynomials in energy in the physical region and then extrapolated to the pole. For a more reliable determination of the ANCs, various forms of functions expressed in terms of phase shifts were used in analytical approximation and subsequent extrapolation. 
\end{abstract}

\maketitle

\section{Introduction}

Asymptotic normalization coefficients (ANC) determine the asymptotics of nuclear wave functions in binary channels at distances between fragments exceeding the radius of the nuclear interaction (see the recent review paper \cite{MBrev} and references therein). In terms of ANCs, the cross sections of peripheral nuclear processes, such as reactions with charged particles at low energies, are parameterized, when, due to the Coulomb barrier, the reactions occur at large distances between fragments. The most important class of such processes is astrophysical 
nuclear reactions occurring in the cores of stars, including the Sun. The important role of ANCs in nuclear astrophysics was first noted in Refs. \cite{Mukh1,Xu}, where it was shown that ANCs determine the overall normalization of cross sections of peripheral radiative  capture reactions (see also Refs. \cite{Mukh2,Mukh3}). 

We note that ANCs are important not only for astrophysics. ANCs turn out to be noticeably more sensitive to theoretical models than such quantities as binding energies or root-mean-square radii. This circumstance makes it possible to use a comparison of the calculated and experimental ANC values to assess the quality of theoretical models. ANCs should be included in the number of important nuclear characteristics along with such quantities as binding energies, probabilities of electromagnetic transitions, etc.

One of the most important astrophysical reactions is the radiative capture of $\alpha$ particles by $^{12}$C. The $^{12}$C$(\alpha,\gamma)^{16}$O reaction is activated during the helium burning stages of stellar evolution. It determines the relative abundance of $^{12}$C and $^{16}$O in the stellar core.  
Although the main contribution to the astrophysical factor of the $^{12}$C$(\alpha,\gamma)^{16}$O process at astrophysial energies comes from two subthreshold bound states $1^{-}$ and $2^{+}$, the radiative capture to 
the excited states $^{16}{\rm O}(0^+)$ and $^{16}{\rm O}(3^-)$ also contributes. Owing to the small binding energies of the considered bound states, the radiative transitions $^{12}{\rm C}(\alpha,\gamma)^{16}{\rm O}(J^{\pi})$ to these states at lower energies relevant the radiative capture  are peripheral. The normalization of the astrophysical $S$-factors for these transitions is determined by the ANCs for the virtual decay $^{16}$O$^*\to \alpha+^{12}$C(g.s.), where g.s. stands for the ground state.
 Hence the  knowledge of these ANCs is important. 
  
	However, 
	{the ANC values available in literature}  for the channels $^{16}{\rm O}^* \to \alpha+^{12}$C(g.s.)  and obtained by various methods are characterized by a noticeable  spread as can be seen from 
	Table \ref{table1} ($\varepsilon$ in this table denotes the binding energy in the channel
	$^{16}$O$(J^\pi)\to \alpha+^{12}$C(g.s.)). The ANC 
	$C_0$ corresponding to $^{16}$O$(0^+)$ was treated in our previous paper \cite{BKMS5}. In the present  paper, we determine the ANCs $C_l$ ($l=3,2,1$)  corresponding to the other three bound excited states of 
	$^{16}$O ($J^\pi=3^-$, $2^+$, $1^-$). As in \cite{BKMS5}, the values of $C_l$ are found  	
	using analytic continuation in the energy plane of the $\alpha^{12}$C partial-wave scattering amplitudes, known from the phase-shift analysis of experimental data. 
 Since we use the analytic continuation, one may consider the obtained values as an experimental ones. 
 
\begin{table*}[htb]
\caption{ANC $C_l$ values in fm$^{-1/2}$ for $^{16}$O$^*(J^\pi)\to \alpha+^{12}$C(g.s.)}
\begin{center}
\begin{tabular}{|c|c|c|c|c|}
\hline
$C_0$; $J^{\pi}=0^+$  & $C_3$; $J^{\pi}=3^-$ & $C_2$; $J^{\pi}=2^+$  & $C_1$; $J^{\pi}=1^-$ & References \\
$\varepsilon=1.113$ MeV & $\varepsilon=1.032$ MeV & $\varepsilon=0.245$ MeV & $\varepsilon=0.045$ MeV & \\
\hline
- & - & (1.11$\pm$0.10)$\times 10^5$ & (2.08$\pm$0.19)$\times 10^{14}$ & \cite{Brune} \\
-                              &  -                             & (1.40$\pm$0.42)$\times 10^{5}$ & (1.87$\pm$0.32) $\times 10^{14}$ & \cite{Balhout} \\
-                              & -                              & (1.44$\pm$0.26)$\times 10^{5}$ & (2.00$\pm$0.69) $\times 10^{14}$ & \cite{Oulevsir} \\
(1.56$\pm$0.09)$\times 10^{3}$ & (1.39$\pm$0.08)$\times 10^{2}$ & (1.22$\pm$0.06)$\times 10^{5}$ & (2.10$\pm$0.14) $\times 10^{14}$ & \cite{Avila} \\
-                              & -                              & 0.213$\times 10^{5}$ & 1.03 $\times 10^{14}$ & \cite{Orlov1} \\
0.4057$\times 10^{3}$           & -                              & 0.505$\times 10^{5}$ & 2.073 $\times 10^{14}$ & \cite{Orlov2} \\
- & -& (1.10-1.31)$\times 10^{5}$ & 2.21(0.07) $\times 10^{14}$ & \cite{Sparen} \\
(0.64-0.74)$\times 10^{3}$ & (1.2-1.5)$\times 10^{2}$ & (0.21-0.24)$\times 10^{5}$ & (1.6-1.9)$\times 10^{14}$ & \cite{Ando} \\
0.293$\times 10^{3}$ & - & - & - & \cite{Orlov3} \\
(0.886-1.139)$\times 10^{3}$ & - & - & - & \cite{BKMS5} \\
- & $(2.17\pm 0.05)\times 10^{2}$ & $(1.42 \pm 0.05)\times 10^5$ & $(2.27 \pm 0.02)\times 10^{14}$ & present \\
\hline 
\end{tabular}
\end{center}
\label{table1}
\end{table*}

The values of ANCs $C_l$ are determined by analytical continuation in center of mass (c.m.) energy $E$ of the  partial-wave amplitudes $f_l(E)$ of elastic scattering of alpha particles on $^{12}$C to the points corresponding to the excited  $^{16}$O$(J^\pi)$ bound states  and lying in the unphysical region of negative  values of $E$. Information on $f_l(E)$ at $E>0$ is taken from the phase-shift analysis.  The obtained ANC values are compared with the results of other authors.

The paper is organized as follows. Section II presents the general formalism of the method used. Section III is devoted to determining $C_3$, $C_1$, and $C_2$ by analytic continuation of experimental data.   
In Section IV, a new rigorous method for the analytic continuation of partial scattering amplitudes is proposed. The results obtained are briefly discussed in Section V.
We use the system of units in which $\hbar=c=$1 throughout the paper.

\section{Basic formalism}

In this section we recapitulate basic formulas which are necessary for the subsequent discussion. 

The Coulomb-nuclear  amplitude of elastic scattering of particles 1 and 2 is of the form
\begin{equation}\label{fNC}
f_{NC}({\rm {\bf  k}})=\sum_{l=0}^\infty(2l+1)\exp(2i\sigma_l)\frac{\exp(2i\delta_l)-1}{2ik}P_l(\cos\theta).
\end{equation}
Here ${\rm {\bf k}}$ is the relative momentum of particles 1 and 2, $\theta$ is the c.m. scattering angle,   
$\sigma_l=\arg\,\Gamma(l+1+i\eta)$  
 and $\delta_l$ are the pure Coulomb and Coulomb-nuclear phase shifts, respectively, $\Gamma(z)$ is the Gamma function, 
\begin{equation}\label{eta}
\eta =Z_1Z_2e^2\mu/k
\end{equation}
is the Coulomb  parameter for the 1+2 scattering state with the relative momentum $k$ related to the energy by  $k=\sqrt{2\mu E}$,
$\mu=m_1m_2/(m_1+m_2)$, $m_i$ and $Z_ie$  are the mass and the electric charge of particle $i$.

The behavior of the Coulomb-nuclear partial-wave amplitude $f_l=(\exp(2i\delta_l)-1)/2ik$ is irregular near 
$E=0$. Therefore, one has to introduce the renormalized Coulomb-nuclear partial-wave amplitude $\tilde f_l$ \cite{Hamilton,BMS,Konig}
\begin{equation}\label{renorm}
\tilde f_l=\exp(2i\sigma_l)\,\frac{\exp(2i\delta_l)-1}{2ik}\,\left[\frac{l!}{\Gamma(l+1+i\eta)}\right]^2e^{\pi\eta}.
\end{equation}
Eq.~(\ref{renorm}) can be rewritten as 
\begin{equation}\label{renorm1}
\tilde f_l=\frac{\exp(2i\delta_l)-1}{2ik}C_l^{-2}(\eta),
\end{equation}
where $C_l(\eta)$ is the Coulomb penetration factor (or Gamow factor) determined by
\begin{align}\label{C}
C_l(\eta)&=\left[\frac{2\pi\eta}{\exp(2\pi\eta)-1}v_l(\eta)\right]^{1/2}, \\ 
v_l(\eta)&=\prod_{n=1}^{l}(1+\eta^2/n^2)\;(l>0),\quad v_0(\eta)=1.
\end{align}
It was shown in Ref. \cite{Hamilton} that  the
analytic properties of ${\tilde f}_{l}$ on the physical sheet of $E$  are analogous to the ones of the partial-wave scattering amplitude for the short-range potential and ${\tilde f}_{l}$  can be analytically continued into the negative-energy region.

The amplitude $\tilde f_l$ can be expressed in terms of the Coulomb-modified effective-range function (ERF) $K_l(E)$ \cite{Hamilton, Konig}  as
\begin{align} 
\label{fK}
\tilde f_l&=\frac{k^{2l}}{K_l(E)-2\eta k^{2l+1}h(\eta)v_l(\eta)}\\ 
&=\frac{k^{2l}}{k^{2l+1}C_l^2(\eta)(\cot\delta_l-i)} \\ 
&=\frac{k^{2l}}{v_l(\eta) k^{2l}\Delta_l(E)-ik^{2l+1}C_l^2(\eta)},
\label{fK3}
\end{align}   
where
\begin{align}\label{scatfun}
K_l(E)&= k^{2l+1} \left[ C_l^2(\eta)(\cot\delta_l-i) + 2 \eta h(k)v_l(\eta) \right],\\ 
h(\eta) &= \psi(i\eta) + \frac{1}{2i\eta}-\ln(i\eta), \\  
\Delta_l(E)&=kC_0^2(\eta)\cot\delta_l, 
\label{Deltal}
\end{align}
$\psi(x)$ is the digamma function and $\Delta_l(E)$ is the $\Delta$ function introduced in Ref. \cite{Sparen}. 

If the $1+2$ system has in the partial wave $l$ the bound state 3 with the binding energy $\varepsilon=\varkappa^2/2\mu>0$, then the amplitude $\tilde f_l$ has a pole at $E=-\varepsilon$. The residue of $\tilde f_l$ at this point is expressed in terms of the ANC
$C^{(l)}_{3\to 1+2}$ \cite{BMS} as
\begin{align}\label{res2}
{\rm res}\tilde f_l(E)|_{E=-\varepsilon}&=\lim_{\substack{E\to -\varepsilon}}[(E+\varepsilon)\tilde f_l(E)] \\
&=
-\frac{1}{2\mu}\left[\frac{l!}{\Gamma(l+1+\eta_b)}\right]^2 \left[C^{(l)}_{3\to 1+2}\right]^2,
\label{res22}
\end{align}
where $\eta_b=Z_1Z_2e^2\mu/\varkappa$ is the Coulomb  parameter for the bound state 3.

In the present paper, as in \cite{BKMS5}, as an object of analytic continuation, we use the function
$\tilde\Delta_l(E)=v_l(\eta)k^{2l}\Delta_l(E)$ ($\Delta$-method \cite{Sparen}). Within this method, the real part of the denominator of the amplitude 
$\tilde f_l(E)$, which for $E > 0$ coincides with $\tilde\Delta_l(E)$ (see (9)), is analytically  approximated at $E>0$ and continued to the region $E < 0$. The amplitude pole condition is formulated as $\tilde\Delta_l^{appr}(-\varepsilon)=0$, where $\tilde\Delta_l^{appr}(E)$ is a function approximating 
$\tilde\Delta_l(E)$ at $E>0$. In practice, for the continuation, it turns out to be more convenient to use not $\tilde\Delta_l(E)$ itself, but some functions that contain it. Some remarks regarding the use of the $\Delta$-method are given in Section IV.

The functions we are considering, determined by the experimental data, are approximated in the physical region $E>0$ by the expression 
\begin{equation}\label{polin} 
\sum_{i=0}^Nc_i P_i(E), 
\end{equation}
where $P_i$ are the Chebyshev polynomials of degree $i$. The maximum degree of the polynomial $N$ and the coefficients $c_i$ are determined from the best description of the approximated functions using the $\chi^2$ criterion and also the F-criterion (see the monograph \cite{Wolberg}). The F-criterion allows us to estimate the probability of that the decrease in the standard deviation when adding the next term to the approximating series really improves the quality of the approximation, and is not random. Note that these two criteria give similar results.

\section{Finding $C_3$, $C_1$, and $C_2$ by analytical continuation of experimental data}

ANCs $C_l$ are found by continuing to the pole $E=-\varepsilon$ phase shifts $\delta_l(E)$ obtained from the phase-shift analysis of the elastic $\alpha-^{12}$C scattering data of Ref. \cite{Tischhauser}. Note that the $\delta_l(E)$ values in \cite{Tischhauser} contain a random error of 5\%. 
   For fitting, 20 points are used for the laboratory energy $E_\alpha$ in the range 2.61 - 6.20 MeV.
Note that near $E=0$,  $\tilde\Delta_l(E)$ changes exponentially which makes it difficult to accurately approximate it by polynomials. Our preliminary calculations showed that direct approximation of $\tilde\Delta_l(E)$ by polynomials leads to poor convergence of results for $C_l$ with increasing degree $N$ of the approximating polynomial. Moreover, some values of $N$ lead to unphysical imaginary values of $C_l$.  Therefore, as in the previous work \cite{BKMS5}, we use the logarithm procedure. Using the logarithmic function makes it possible to soften this exponential dependence and improve the quality of approximation of the considered functions. 

\subsection{ANC $C_3$ for $3^-$ state of $^{16}$O}

For this state, the binding energy in the $\alpha+^{12}$C(g.s.) channel is $\varepsilon =1.032$ MeV.
  As follows from \cite{Tischhauser}, the function $\tilde\Delta_3(E)$ changes rapidly in the region where it is approximated, in particular, $\tilde\Delta_3(E)=0$ at $E=E_z=4.41$ MeV, that is, at  $E_\alpha=5.89$ MeV. Taking this circumstance into account, the function to be approximated was chosen in the form
	\begin{equation} \label{F3}
F_3(E) = \ln\left(A-\dfrac{\tilde\Delta_3(E)}{E-E_z}\right).
\end{equation}
Equation (\ref{F3}) differs from the form of the function used to approximate the experimental data in \cite{BKMS5} only by the presence of the factor $(E-E_z)^{-1}$ which ensures the constancy of the sign of the quantity $\tilde\Delta_3(E)/(E-E_z)$ in the entire region where it is approximated.

The constant $A>0$ is added to make $A-\tilde\Delta_3(-\varepsilon)/(-\varepsilon-E_z)$ positive.  As in \cite{BKMS5}, the value of $A$ is chosen so that within the approximating energy range the condition 
$A \ll |\Delta_0(E)|$ holds, and the approximated function is as close to a straight line as possible so that it could be approximated by a polynomial of a low degree $N$. 
	In practice, the value of $A$ was found from the requirement that on the curve of the energy dependence of the approximated function, three points located near $E = 0$ lay on a straight line.  $E=-\varepsilon$ was chosen as one of these points, the other two points were taken at $E>0$. 
As will be seen below, when these requirements are met, the calculated ANC values weakly depend on the $A$ values.

To determine the sensitivity of the results to parameter $A$,  calculations for all considered $l$ values  have been performed for two different $A$ values: $A_1$ and $A_2$.
	The value of $A_1$  is derived from the procedure described above using the phase shift 
$\delta_l(E_\alpha)$ values at $E_\alpha=2.61$ MeV and $E_\alpha=4.12$ MeV. $A_2$ corresponds to the choice 
$E_\alpha=2.61$ MeV and $E_\alpha=6.01$ MeV. For $l=3$ $A_1$=0.0210 fm$^{-6}$ and $A_2$= 0.0241 fm$^{-6}$.	
	
	The results of calculations of ANC $C_3$ are presented in Table \ref{table2}. The column  labeled ``F-criterion'' in Table \ref{table2} and in subsequent tables   gives the probability that adding the next term to the approximation series leads to an improvement in the approximation. The closer the value is to 100\%, the more justified is the addition of the next term \cite{Wolberg}. 
From Table \ref{table2} it follows that the combined use of both criteria ($\chi^2$ and F) select 
$C_3=215$ fm$^{-1/2}$ for $A_1$ and $C_3=212$ fm$^{-1/2}$ for $A_2$ as the best results.  

\begin{table*}[htb]
\caption{ANC $C_3$ for $J^\pi=3^-$}
\begin{center}
\begin{tabular}{|c|c|c|c|c|c|c|}
\hline
& \multicolumn{3}{c|}{$A_1$} & \multicolumn{3}{c|}{$A_2$} \\
\hline
$N$  & $C_3$, fm$^{-1/2}$  & $\chi^2$  & F-criterion, \% & $C$, fm$^{-1/2}$  & $\chi^2$  & F-criterion, \% \\
\hline
1 & 234 & 0.527 & $>$99\% & 224 & 0.551 & $>$99\%\\
2 & 231 & 0.294 &  95\% & 227 & 0.276 &  92\%\\
3 & 215 & 0.244 &  44\% & 212 & 0.243 &  41\%\\
4 & 200 & 0.253 &   7\% & 197 & 0.253 &   6\%\\
5 & 188 & 0.270 &  29\% & 185 & 0.270 &  29\%\\
\hline
\end{tabular}
\end{center}
\label{table2}
\end{table*}

To improve the reliability of the determination of $C_3$, calculations based on Eq. (\ref{F3}) were also carried out using 11 experimental points lying in a narrower energy interval
(up to $E_\alpha=4.31$ MeV).
	The results of calculations using $A=A_1$ are presented in Table \ref{table3}. We choose $C_3=225$ 
	fm$^{-1/2}$ as the best result. 
As a final result, we take the average of the three obtained values  (two values from Table II and one from Table III): $C_3=(217\pm 5)$ fm$^{-1/2}$.

\begin{table}[htb]
\caption{ANC $C_3$ for $J^\pi=3^-$. Narrow energy range}
\begin{center}
\begin{tabular}{|c|c|c|c|}
\hline
$N$  & $C_3$, fm$^{-1/2}$  & $\chi^2$  & F-criterion\\
\hline
1 & 233 & 0.390 & 94\% \\
2 & 225 & 0.272 & 27\% \\
3 & 214 & 0.306 & 68\% \\
4 & 14.3  & 0.297 & 3\% \\
5 & 156 & 0.356 & 12\% \\
\hline
\end{tabular}
\end{center}
\label{table3}
\end{table}

\begin{figure}[htb]
\includegraphics[scale=0.95]{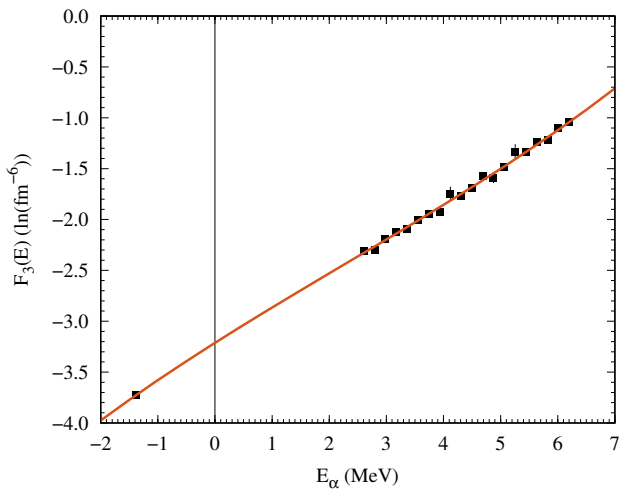} 
\caption {The approximated function $F_3(E)$. $A=A_2$, $N=3$.  Experimental points are taken from Ref. \cite{Tischhauser}}
\label{fig1}
\end{figure}

\begin{figure}[htb]
\includegraphics[scale=0.95]{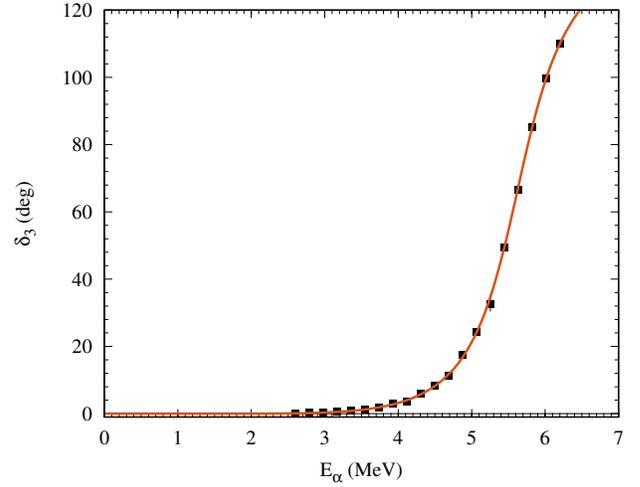} 
\caption {Approximated  $\delta_3(E)$. $A=A_2$, $N=3$.  Experimental points are taken from Ref. \cite{Tischhauser}}
\label{fig2}
\end{figure}

Comparison of $F_3(E)$ and $\delta_3(E)$  obtained by fitting based on Eq. (\ref{F3}) for $A=A_2$ with phase=shift analysis data is shown in Figs. \ref{fig1} and \ref{fig2}.

\subsection{ANC $C_1$ for $1^-$ state of $^{16}$O}

For this state, the binding energy in the $\alpha+^{12}$C(g.s.) channel is $\varepsilon =0.045$ MeV.
The calculations for $C_1$ were carried out similarly to the calculations for $C_3$ described in the previous subsection. They were based on  Eq. (\ref{F3}) but with index 3 replaced by 1. Experimental data were taken from the same paper \cite{Tischhauser}. The same energy ranges were used for the analysis. The values of the constants $A_1$ and $A_2$ were determined using the procedure described in subsection 3.1 what results in  $A_1$=0.0136 fm$^{-2}$ and $A_2$= 0.0362 fm$^{-2}$. Phase shift $\delta_1(E)$ vanishes at $E=E_z=2.43$ MeV.

The results of $C_1$ calculations are presented in Tables \ref{table4} and \ref{table5}. 

\begin{table}[htb]
\caption{ANC $C_3$ for $J^\pi=1^-$}
\begin{center}
\begin{tabular}{|c|c|c|c|c|c|c|}
\hline
& \multicolumn{3}{c|}{$A_1$} & \multicolumn{3}{c|}{$A_2$} \\
\hline
$N$  & $C_1$, fm$^{-1/2}$  & $\chi^2$  & F-criterion & $C_1$, fm$^{-1/2}$  & $\chi^2$  & F-criterion \\
\hline
1& 2.49$\times 10^{14}$ & 18.1 & $>$99\% & 2.06$\times 10^{14}$ & 1.22 & $>$99\%\\
2& 2.19$\times 10^{14}$ &  0.320 &  97\% & 2.00$\times 10^{14}$ & 0.761 & $>$99\%\\
3& 2.27$\times 10^{14}$ &  0.255 &  31\% & 2.31$\times 10^{14}$ & 0.255 &  26\%\\
4& 2.35$\times 10^{14}$ &  0.268 &  70\% & 2.43$\times 10^{14}$ & 0.269 &  68\%\\
5& 1.62$\times 10^{14}$ &  0.265 &  70\% & 1.37$\times 10^{14}$ & 0.268 &  68\%\\
\hline
\end{tabular}
\end{center}
\label{table4}
\end{table}

\begin{table}[htb]
\caption{ANC $C_1$ for $J^\pi=1^-$. Narrow energy range}
\begin{center}
\begin{tabular}{|c|c|c|c|}
\hline
$N$  & $C_1$, fm$^{-1/2}$  & $\chi^2$  & F-criterion\\
\hline
1 & 2.40$\times 10^{14}$ & 0.56834 & $>$99\% \\
2 & 2.24$\times 10^{14}$ & 0.25940 &  49\% \\
3 & 2.47$\times 10^{14}$ & 0.27777 &  52\% \\
4 & 1.43$\times 10^{14}$ & 0.29582 &  25\% \\
5 & 7.31$\times 10^{14}$ & 0.34724 &  52\% \\
\hline
\end{tabular}
\end{center}
\label{table5}
\end{table}
For a wider energy range (Table \ref{table4}), the criteria $\chi^2$ and F select $C_1=2.27\times 10^{14}$ 
fm$^{-1/2}$ ($A=A_1$, $N=3$) and $C_1=2.31\times 10^{14}$ fm$^{-1/2}$ ($A=A_2$, $N=3$). Table \ref{table5} results in $C_1=2.24\times 10^{14}$ fm$^{-1/2}$ ($N=2$). 
For the mean value of the ANC, we obtain $C_1=(2.27\pm 0.02)\times 10^{14}$ fm$^{-1/2}$. 

\begin{figure}[htb]
\includegraphics[scale=0.95]{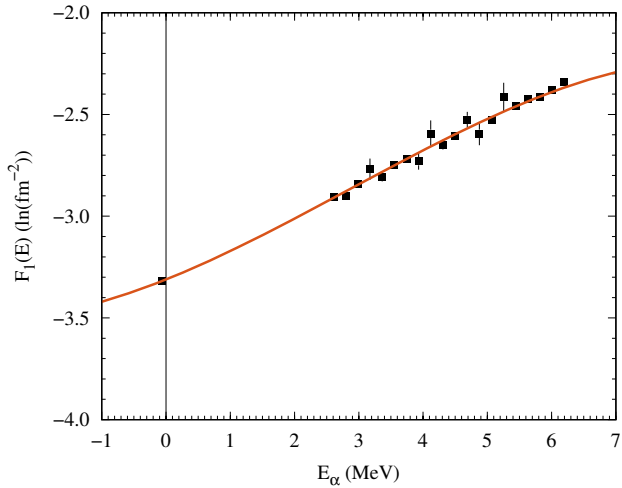} 
\caption {The same as in Fig. 1 but for $F_1(E)$.  $A=A_2$, $N=3$}
\label{fig3}
\end{figure}

\begin{figure}[htb]
\includegraphics[scale=0.95]{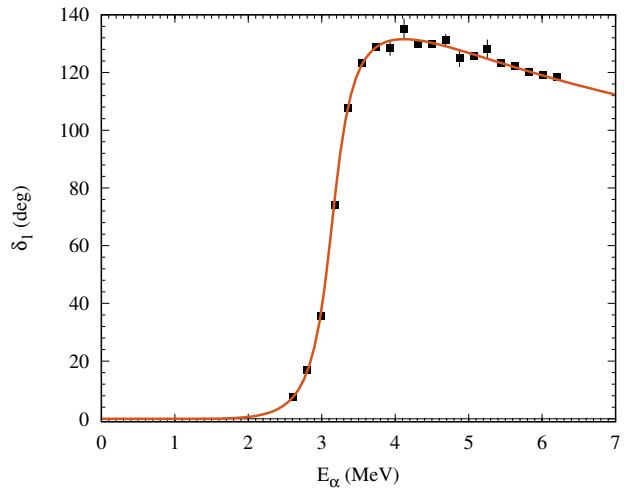} 
\caption {The same as in Fig. 2 but for $\delta_1(E)$. $A=A_2$, $N=3$}
\label{fig4}
\end{figure}

Comparison of $F_1(E)$ and $\delta_1(E)$  obtained by fitting based on Eq. (\ref{F3}) for $A=A_2$ with phase-shift analysis data is shown in Figs. \ref{fig3} and \ref{fig4}.

\subsection{ANC $C_2$ for $2^+$ state of $^{16}$O}

For this state, the binding energy in the $\alpha+^{12}$C(g.s.) channel is $\varepsilon =0.245$ MeV.
To determine $C_2$, we used phase shift $\delta_2$ data from the same work \cite{Tischhauser} and the same energy ranges as in subsections 3.1 and 3.2. 
However, $\delta_2(E)$ has a very complex behavior at low energies. As a result, $\tilde\Delta_2(E)$ in the energy range of interest to us has two poles at $E=E_{p1}$=2.67 MeV and $E=E_{p2}$=3.98 MeV and two zeros   at $E=E_{z1}$=2.68 MeV and $E=E_{z2}$=4.36 MeV. It is practically impossible to approximate such a complex behavior directly. In work \cite{Ando}, to find $C_2$, a very narrow energy interval ($E=1.95-2.4$ MeV) was used, which does not include poles and zeros of $\tilde\Delta_2(E)$. In the work \cite{Orlov2}, the poles were eliminated by multiplying $\tilde\Delta_2(E)$ by $(1-E/E_{p1})(E-E_{p2})$. However, due to the presence of two zeros, the resulting function was quite complex and difficult to accurately approximate.

In the present paper, we have used the function
\begin{equation} \label{G}
G(E)=\tilde\Delta_2(E) \dfrac{(E-E_{p1})(E-E_{p2})}{(E-E_{z1})(E-E_{z2})}.
\end{equation}
Function $G(E)$ in the energy range of interest to us contains neither poles nor zeros and smoothly depends on the energy.
Specifically, by analogy with the cases $l=3$ and $l=1$, for polynomial approximation and subsequent analytic continuation to a point $E=-\varepsilon$, the following function was used:
\begin{equation} \label{F2}
F_2(E)=\ln[A-G(E)].
\end{equation}
Calculations for $C_3$ were carried out similarly to the calculations for $C_3$ and $C_1$.  They were based on  Eqs. (\ref{G}) and (\ref{F2}). For constants $A_1$ and $A_2$, the procedure described in the subsection 3.1. leads to values $A_1=0.354\times 10^{-4}$ fm$^{-5}$ and $A_2= 0.461\times 10^{-4}$ fm$^{-5}$.

The results of $C_2$ calculations are presented in Tables \ref{table6} and \ref{table7}. 

\begin{table}[htb]
\caption{ANC $C_2$ for $J^\pi=2^+$}
\begin{center}
\begin{tabular}{|c|c|c|c|c|c|c|}
\hline
& \multicolumn{3}{c|}{$A_1$} & \multicolumn{3}{c|}{$A_2$} \\
\hline
$N$  & $C_2$, fm$^{-1/2}$  & $\chi^2$  & F-criterion & $C_2$, fm$^{-1/2}$  & $\chi^2$  & F-criterion \\
\hline
1 & 1.49$\times 10^5$ & 19.2 & $>$99\% & 1.35$\times 10^5$ & 13.6 & 46\%\\
2 & 1.46$\times 10^5$ & 13.3 &  49\% & 1.35$\times 10^5$ & 14.1 & 74\%\\
3 & 1.50$\times 10^5$ & 13.7 &  31\% & 1.434$\times 10^5$ & 13.8 & 35\%\\
4 & 1.75$\times 10^5$ & 14.4 &  61\% & 1.79$\times 10^5$ & 14.5 & 61\%\\
5 & 0.803$\times 10^5$ & 14.7 &  70\% & 0.726$\times 10^5$ & 14.7 & 70\%\\
\hline
\end{tabular}
\end{center}
\label{table6}
\end{table}

\begin{table}[htb]
\caption{ANC $C_2$ for $J^\pi=2^+$. Narrow energy range}
\begin{center}
\begin{tabular}{|c|c|c|c|}
\hline
$N$  & $C_2$, fm$^{-1/2}$  & $\chi^2$  & F-criterion\\
\hline
1 & 1.45$\times 10^5$ & 25.3 & 42\% \\
2 & 1.56$\times 10^5$ & 27.3 & 51\% \\
3 & 1.13$\times 10^5$ & 29.0 & 62\% \\
4 &  0.491$\times 10^5$ & 29.4 & 89\% \\
5 & -      & 19.9 & 66\% \\
\hline
\end{tabular}
\end{center}
\label{table7}
\end{table}

Using the criteria $\chi^2$ and F, we select from Table \ref{table6} $C_2=1.46\times 10^5$ fm$^{-1/2}$ ($A=A_1$, $N=2$) and $C_2=1.35\times 10^5$ fm$^{-1/2}$ ($A=A_2$, $N=1$), and from Table \ref{table7} $C_2=1.45\times 10^5$ fm$^{-1/2}$ ($N=1$). A dash in Table \ref{table7} means that the corresponding variant
leads to an unphysical imaginary value of $C_2$.
For the mean value, we get $C_2=(1.42\pm 0.05)\times 10^5$ fm$^{-1/2}$.

\begin{figure}[htb]
\includegraphics[scale=0.95]{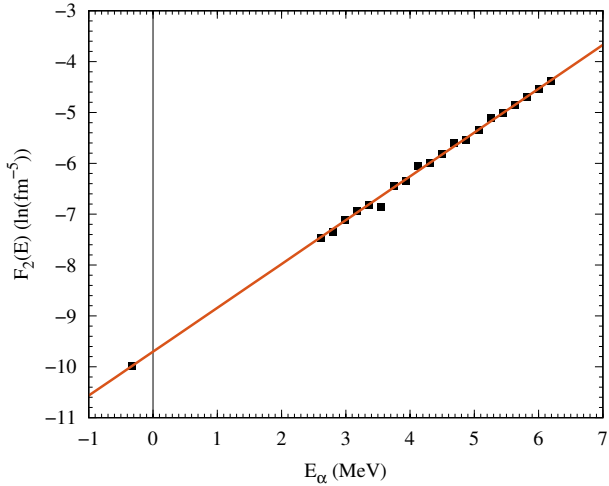} 
\caption {The same as in Fig. 1 but for $F_2(E)$. $A=A_2$, $N=1$}
\label{fig5}
\end{figure}

\begin{figure}[htb]
\includegraphics[scale=0.95]{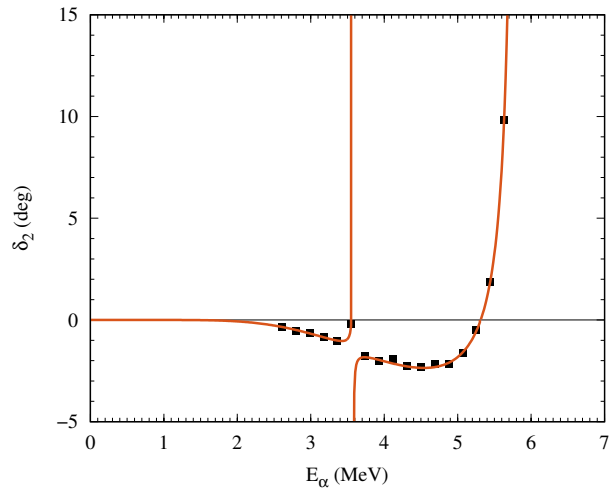} 
\caption {The same as in Fig. 2 but for $\delta_2(E)$. $A=A_2$, $N=1$}
\label{fig6}
\end{figure}

Comparison of $F_2(E)$ and $\delta_2(E)$  obtained by fitting based on Eqs. (\ref{G}) and (\ref{F2}) for 
$A=A_2$ with phase-shift analysis data is shown in Figs. \ref{fig5} and \ref{fig6}.

\section{A possible new method of analytic continuation of experimental data}

 In connection with the use of the $\Delta$-method in the present work, it should be reminded that, according to the conclusions of Refs. \cite{BKMS2,Gaspard} this method can be employed to obtain information on bound states if their energy and the 
energy of scattering states used to approximate the $\Delta$ function satisfy the condition
  \begin{equation}\label{range}
|E|\le (Z_1Z_2e^2)^2\mu/2=1\,\mathrm{Ry},
\end{equation}
where 1 Ry is the nuclear Rydberg energy. 
For  the $\alpha+^{12}$C system  1 Ry = 10.7 MeV and the condition \eqref{range} is fulfilled in the present work.  However, in general, this method of analytic continuation of scattering data is not quite strict and correct from the point of view of mathematics. Note that for lighter systems, in particular for the channels $^6$Li$\to \alpha+d$ and  
$^7$Be$\to \alpha+^3$He,  the $\Delta$-method is not suitable  due to a very narrow range of allowed energy values.  

 On the other hand, the method based on the continuation of the effective range function $K_l(E)$ \eqref{fK}, although formally rigorous, is practically suitable only for the lightest nuclear systems due to the presence of a large background of purely Coulomb terms. In particular, for the $\alpha+^{12}$C system  considered in this paper, any reliable continuation of  
$K_l(E)$ to the region $E<0$, taking into account experimental errors, turned out to be impossible.

 In this regard, we would like to point out a possible alternative method of analytic continuation devoid of the above disadvantages.
Let us write $\tilde f_l(E)$ \eqref{fK} in the form
\begin{align}\label{D}
\tilde f_l(E)&=\frac{k^{2l}}{D_l(E)},\\
D_l(E)&=\tilde \Delta_l(E)-ik^{2l+1}C_l^2(\eta)\\
&=K_l(E)-2\eta k^{2l+1} h(\eta) v_l(\eta).
\end{align}
Introduce the quantity $g_l(E)$ according to
\begin{align} \label{g}
g_l(E)&= D_l(E)-D_l^{SW}(E)\\
&=\tilde\Delta_l(E)-\tilde\Delta_l^{SW}(E)\\
&=K_l(E)-K_l^{SW}(E),
\end{align}
where $D_l^{SW}(E)$, $\tilde\Delta_l^{SW}(E)$ and $K_l^{SW}(E)$ are the functions $D_l(E)$, $\tilde\Delta_l(E)$ and $K_l(E)$ for a potential that is the sum of the Coulomb and square well (SW) potentials. $\tilde\Delta_l^{SW}(E)$ is expressed explicitly in terms of parameters of a SW potential (see Eq. (16) from Ref. \cite{BKMS5}).
$K_l(E)$ and $K_l^{SW}(E)$, as distinct from $\tilde\Delta_l(E)$, have no essential singularity at $E=0$ and can be expanded in a series in $E$ near $E=0$. Therefore, the function $g_l(E)$ can also be expanded into a series in $E$ near $E=0$. We emphasize that this property takes place for any parameters of the SW potential.  Hence, $g_l(E)$ can be approximated by a polynomial at $E>0$ and continued to the negative energy region. Note  that $g_l(E)$ is real for both positive and negative energies.

The condition of the pole of $\tilde f_l(E)$ at $E=-\varepsilon=-\varkappa^2/2\mu$ is
\begin{equation}\label{cond}
D_l(-\varepsilon)=g_l(-\varepsilon)+D_l^{SW}(-\varepsilon)=0. 
\end{equation}
	ANC $C_l$ is defined by the expression
	\begin{align} \label{Cnew}
C_l^2&=-2\mu\left[\frac{\Gamma(l+1+\eta_b)}{l!}\right]^2
{\rm res} \tilde f_l(E)|_{E=-\varepsilon}\\
&=    
-2\mu\left[\frac{\Gamma(l+1+\eta_b)}{l!}\right]^2(-1)^l\varkappa^{2l}\left(\frac{dD_l(E)}{dE}\right)^{-1}_{E=-\varepsilon}.
\end{align}
The proposed approach is completely rigorous. The term $D_l^{SW}(E)$ plays the role of a background; however, it does not contain large pure Coulomb contributions. Morover, by varying the parameters of the SW potential, it can be made small compared to $D_l(E)$ in the energy region of approximation of the experimental data. 

\section{Conclusions}

In the present paper, we treated the ANC $C_l$ corresponding to the virtual decay of three excited bound  states of $^{16}$O$(J^\pi)$ ($J^\pi=3^-,2^+,1^-;\,l=J$) to $\alpha+^{12}$C(g.s.). The ANC $C_0$ for the excited state $^{16}$O$(0^+; 6.05$ MeV) was considered in our previous work \cite{BKMS5}. The  values of $C_l$ obtained by various methods and presented in Table \ref{table1} are characterized by a large spread. As for the ground state of $^{16}$O, it is hardly possible to determine the corresponding ANC $\bar C_0$  by analytic continuation of the data on partial-wave scattering amplitudes (see Refs. \cite{BKMS2,BlSav2016}). $\bar C_0$ values obtained by other methods can be found in Ref. \cite{Shen}. 

To determine $C_l$, we use  analytic continuation in energy of experimental $\alpha-^{12}$C scattering data to the poles of the partial-wave scattering amplitudes corresponding to bound states of $^{16}$O$(J^\pi$).  
Specifically, the analytic continuation was carried out on the basis of polynomial approximation  and subsequent extrapolation of some expressions containing the function $\tilde\Delta_l(E)$ defined above. 
$\tilde\Delta_l(E)$ is expressed in terms of phase shifts.

The mean ANC values obtained by averaging the results of various approximation options are presented in the last line of Table \ref{table1}.
Comparing our results with the values obtained earlier (see Table \ref{table1}), we see that the value of 
$C_3$ found by us exceeds the previous results. As for the ANC $C_2$, the values presented in 
Table \ref{table1} are characterized by a very large spread. The value we obtained is close to the maximum values from Table \ref{table1}. Finally, our $C_1$ value is close to most of the previously obtained results, although it slightly exceeds them. 

We plan to test the efficiency of the new rigorous method described in Section IV for extending the scattering data to the region of negative energies. 
		
\section*{Acknowledgements}

 A.S.K. acknowledges the support from the Australian Research Council. A.M.M. acknowledges the support from the US DOE National Nuclear Security Administration under Award Number DENA0003841 and DOE Grant No. DE-FG02-93ER40773.

\end{document}